\documentclass[aps,prl,twocolumn,groupedaddress,amsmath]{revtex4}

\usepackage{graphicx}

\begin{document}




\title{Universal Laws and Economic Phenomena}



\author{Austin Gerig}
\email{austin.gerig@uts.edu.au}
\affiliation{School of Finance and Economics, University of Technology, Sydney, Australia}



\begin{abstract}

\end{abstract}



\maketitle


\begin{quote}
\emph{``The supreme task of the physicist is to arrive at those universal elementary laws from which the cosmos can be built up by pure deduction.'' Albert Einstein, 1918.}
\end{quote}

When researching the natural world, scientists often search for universal principles or laws to explain the systematic working of things.  This approach has served them well, but can it be applied to disciplines outside the natural sciences?  Are there universal laws, for example, that underlie social and economic phenomena, and should economists search for such laws?  I believe the answer is yes, and that it is advantageous for economists to adopt methods from the natural sciences to uncover them.

In the natural sciences, universal laws are discovered through data and experimentation, often by searching for regularities that hold under a wide range of circumstances.   This approach is rarely used in economics.  Instead, economic theories are often based on assumptions about how the world \emph{should} be and how agents \emph{should} behave, e.g., agents are assumed to be perfectly rational and markets are considered complete and in equilibrium\cite{Farmer08}.  This means that in much of economic theory, the universal principles that underlie phenomena are agreed upon \emph{a priori}, rather than discovered through empirical analysis.  It also means that theories are often accepted and advanced based on mathematical elegance or rigor rather than correspondence with data. 

This method may seem strange to scientists, but there are several good reasons why economists have adopted it.  First, economic systems are mutable.  We are part of economic systems after all, and unlike particles, we can change our behavior or the rules of the game to correspond with how things should be.  Second, there is something compelling about thinking that economic systems are okay---that they are in the hands of rational agents and have a robust structure such that things are stable.  Finally, these theories serve as reference points to understand the world.  It's interesting to know if and when economic phenomena depart from these theories because it suggests we have work to do.

Complementary to this methodology, however, can be the methodology of the natural sciences, where emphasis is placed on finding regularities in data, and where the underlying cause is expressed in the form of some universal principle or law that is repeatedly tested.  Sometimes the underlying cause will correspond with what we originally thought it should be, but sometimes it may not.  To illustrate what I mean, consider the way that prices move in financial markets.

\section{Random Walks}

Over a century ago, the French mathematician Louis Bachelier proposed that stock prices follow a \emph{random walk}---that prices move up or down in random increments such that price changes are unpredictable\cite{Bachelier64}.  When analyzing economic data, the random walk model is surprisingly accurate.  It holds not only for stock prices, but also for the prices of many other items: stock indices, derivative instruments, commodities and other economic goods, and even for the prices of contracts traded on prediction markets.  The regularity of this behavior across different items hints that some fundamental mechanism is behind it;  perhaps some universal principle is at work.  

In fact, most economists believe this is true, and they attribute the randomness of prices to the profit maximization (or loss aversion) of investors.  If prices did not move randomly, but instead were in some way predictable, then this predictability would be quickly removed.  After all, who would be willing to sell a stock for $\$90$ if everyone knew the price would move up to $\$100$ during the next period?  Wouldn't sellers try to get something closer to $\$100$ right now, and wouldn't buyers be willing to pay something closer to $\$100$?  When these individuals push the price to $\$100$, the predictability in the price movement is removed.  If predictable price movements quickly disappear, then the only way for prices to move is with random increments.  Paul Samuelson, an American economist, derived this result in his paper entitled ``Proof That Properly Anticipated Prices Fluctuate Randomly''\cite{Samuelson65}.  It is an elegant and simple explanation for the universally observed random nature of price movements.  The theory explains a large collection of phenomena and has predictive power (it predicts that any prices determined by profit maximizing agents will be random)---both are hallmarks of theories developed by seeking universal laws.

\begin{figure*}[htb]
\centering
\includegraphics[width=3.4in]{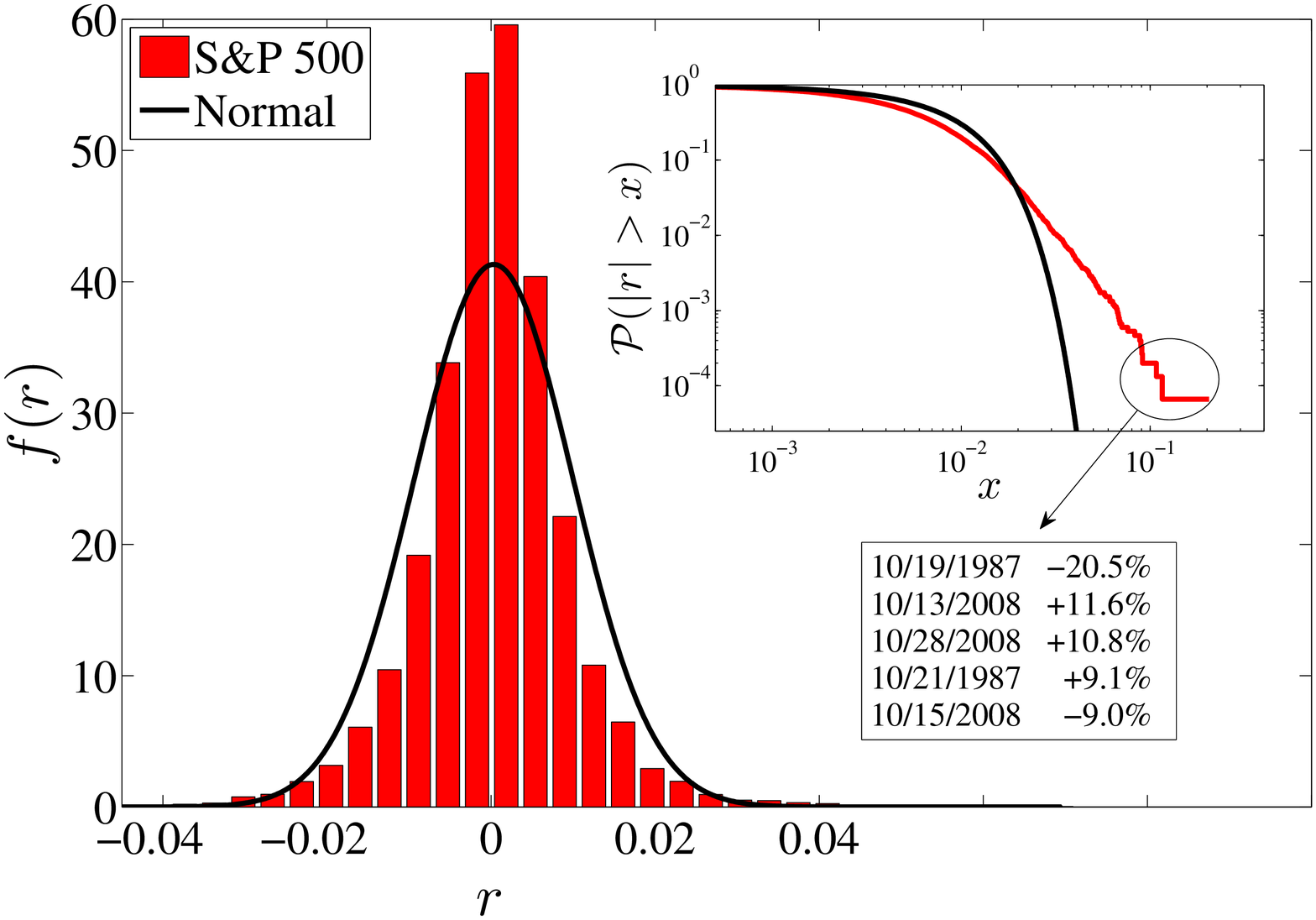}
\includegraphics[width=3.4in]{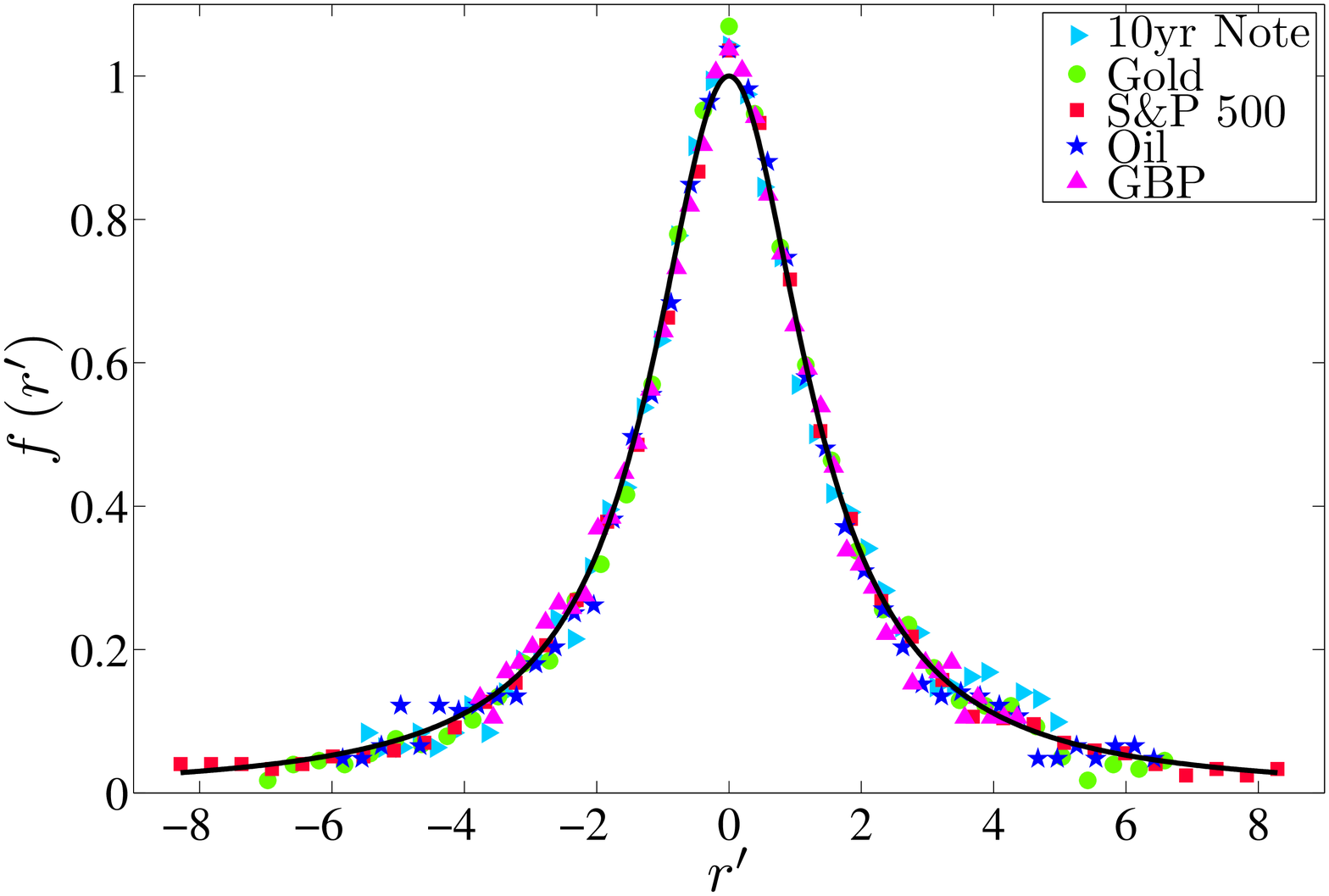}
\caption{(a) The probability distribution of daily returns for the S\&P 500 stock index from January 3, 1950 to November 25, 2009.  Inset: Probability that the absolute daily return is above a threshold value, $x$.  The five largest absolute returns are circled and their dates and sizes are shown.  (b) The rescaled probability distribution of daily returns for six different traded items. Adapted from \cite{Gerig10}.}
\end{figure*}

\section{Extreme Price Movements}

There is another interesting regularity found in economic prices: very large price movements, such as stock market crashes, occur frequently\cite{Mandelbrot63, Fama63, Mandelbrot08}.  This, again, happens across the board for many different economic items.  

To understand just how large these price movements are, consider what it would mean if human heights behaved in a similar way.  Assume for a moment that adult human heights varied between individuals in the same way that price movements vary.  Within your circle of friends, there wouldn't be that much of a difference; most people would be between 5 to 6 ft tall.  Outside of this circle, however, there would be dramatic changes.  In your city, someone would be over 30 ft tall.  In your country, the tallest person would likely reach 150 ft, and the tallest person in the world would be over 1000 ft tall.  
  
The distinction between human heights and price movements is not just a pedagogical exercise, it is important because most financial models assume that the distribution of stock returns is the same as the distribution pattern for human heights -- the ubiquitous bell-shaped curve known as the normal (or Gaussian) distribution.  If this were the case, very large returns (analogous to a 150 foot person) should never occur.  But this is incorrect.  For reasons we do not fully understand, stock returns are not distributed according to a normal distribution.  Instead, they have a much larger peak and the `tails' or extremes of the distribution are thicker.  This means that large price movements occur more often than predicted.

In Fig.~1(a), I show the distribution for the daily returns of the S\&P 500 stock index from January 3, 1950 to November 25, 2009.  This plot can be replicated by downloading data from http://finance.yahoo.com.  The horizontal axis measures the different sizes of returns (0.02 is a $2\%$ return, 0.04 is a $4\%$ return, etc.) and the vertical axis shows the relative likelihood of these price changes -- the higher the red bar, the more likely that event is observed.  Small returns, close to zero, are the most likely occurrence.  A normal distribution is fit to the data and is drawn with a black line.  Notice that this does not coincide well with the S\&P 500 data.

The inset plot shows the probability that a daily return is above a certain threshold value.  It enlarges the tail of the distribution---the area where large price movements are recorded.  You can see that the probability of large returns is much higher than what the normal distribution predicts, i.e., the red curve is above the black curve for large values of $x$.  I've circled the five highest returns and show their values and the dates they were observed.  Not surprisingly, the largest return occurred on Black Monday, October 19, 1987, when stock markets crashed around the world.

If you look at the y-axis in the inset plot, the probability for a daily return to exceed $10\%$ is around $10^{-4}$, which means this has occurred approximately once out of $1/10^{-4}=10,000$ trading days, or once every 40 years.  For comparison, the black curve---a normal distribution---predicts this to occur once every $7\times10^{18}$ years, which is longer than the age of the universe and for all practical purposes, means never.  Obviously this is incorrect.

One way to explain the discrepancy between observed stock returns and financial models is to consider large price movements as outliers---surprising events outside of the normal model.  There are several reasons to do this.  First, there are good underlying reasons to assume a normal distribution for returns as a first guess, and there is no accepted theory for why it should be otherwise.  Second, we usually explain large price movements in this way---stock markets crashed because computer trading malfunctioned or the global financial crises occurred because banks made large mistakes.  When using these explanations, we implicitly suggest that they are one-time events---outliers---that can be accounted for and controlled in the future.  The problem is, despite our efforts, they keep happening.

An alternative explanation is that something more fundamental is producing these events and that the widely reported and agreed-upon culprits are just symptoms of the same underlying cause.  There are several reasons to believe this is true.  First, extreme price movements are not just observed for stocks; these events occur universally across traded items.  Second, the empirical evidence does not show these events as statistical outliers.  You can see this for the S\&P 500 index in the inset plot where the red curve extends continuously in a smooth way down to the points where extreme price movements are recorded.  These points do not exist by themselves but nicely fit where you'd expect them when extrapolating the red curve from smaller price movements.  Finally, there is evidence that the probability distribution of price returns is universal, that it deviates from the normal distribution in the exact same way for different items and over different time periods\cite{Fuentes09,Gerig10}.  In Fig.~1(b), I show the probability distribution for daily returns for five different traded items.  By appropriately rescaling the axes for each, the distributions collapse on the same non-normal curve.  Why would these unrelated price series behave in the same way unless something fundamental was the cause?

Despite the idiosyncratic behavior of individuals, regularities exist in social and economic systems that are similar to those found in natural systems.  Specific examples are found in the way that economic prices behave.  The reason prices are random is well understood; it occurs because individuals are profit maximizing.  The reason prices deviate from a normal distribution is not understood and is currently a matter of much debate.  I believe the evidence suggests some universal mechanism underlies these deviations, and that large price movements are not outliers to an otherwise correct (normal) model.  If true, understanding this mechanism is extremely important.  If large price movements result from human behavior or the way in which markets are structured, then there might be ways to curtail behavior or structure markets differently such that these extreme events do not occur.  If they are due to some economic cause, then perhaps it is something we can only understand and better prepare for.  At least then we would have correct models on which to base economic decisions.

\end{document}